
\normalbaselineskip=12pt
\baselineskip=12pt
\magnification=1200
\hsize 15.2truecm \hoffset 0.5 truecm
\vsize 23.0truecm
\nopagenumbers
\headline={\ifnum \pageno=1 \hfil \else\hss\tenrm\folio\hss\fi}
\pageno=1
\font\fthreeb=cmbx10 scaled\magstep1
\def\lsim{\mathrel{\rlap{\lower4pt\hbox{\hskip1pt$\sim$}}
    \raise1pt\hbox{$<$}}}         
\def\gsim{\mathrel{\rlap{\lower4pt\hbox{\hskip1pt$\sim$}}
    \raise1pt\hbox{$>$}}}         

\centerline{\fthreeb The Crucial Role of $\omega \pi^+$
Production in the $D_s^+$ Decay}

\vskip 60pt
\centerline{ M. Anselmino$^*$, I. Bediaga$^{**}$ and E. Predazzi$^*$}
\vskip 24pt
\centerline{(*) {\it Dipartimento di Fisica Teorica, Universit\`a di Torino}}
\centerline{\it and Istituto Nazionale di Fisica Nucleare,
Sezione di Torino}
\centerline{\it Via P. Giuria 1, I--10125 Torino, Italy}
\vskip 12pt
\centerline{(**) {\it Centro Brasileiro de Pesquisas F\'\i sicas/CNPq}}
\centerline{\it Rua Dr. Xavier Sigaud 150, 22290-180, Rio de Janeiro, Brazil}
\vskip 1.2in
\centerline{\bf ABSTRACT}
\vskip 12pt
\noindent
In this paper the relevance of non-spectator decays of charm
particles is analysed and some crucial tests for it are suggested.
\vskip 1.8 in
\line{\hfil CBPF-NF-45-95}
\line{\hfil DFTT 42/95}
\line{\hfil July 1995}
\line{\hfill hep-ph/9507292}
\vfill
\eject
\baselineskip 18pt plus 2pt minus 2pt

A longstanding (unsolved) problem in charm decay has been
to provide a reliable estimate of non-spectator contributions
in the decay of charm particles (in particular of the so-called
$W$-annihilation). In this paper we are going to analyse a number of
issues related to this problem in the light of a recent analysis of
new data of the E-687 Collaboration [1]:
\item{i)} first, we discuss several points connected with the case of
three-pion $D_s^+$ decay and point out
that some crucial measurements could allow an
uncontroversial assessment of the importance (or of the insignificance)
of non-spectator contributions. Specifically, we suggest that a
precise measurement of the channel $D_s^+ \rightarrow \omega \, \pi^+$
could cast the final word on the r\^ole of the $W$-annihilation
non-spectator component in the decay of $D_s^+$.
\item{ii)} Second, we question the importance of the $s \bar s$
component in $f_0(980)$, $f_0(1300)$ and $f_2(1270)$.
\item{iii)} Next, we propose a specific mechanism that accounts
for the absence of $D_s^+ \rightarrow \rho^0 \pi^+$ (another
longstanding problem in the theoretical analysis of $D_s^+$ decay) and
we suggest ways to check it.
\item{iv)} Finally, should not the decay
$D_s^+ \rightarrow \omega \, \pi^+$
be observed with a significant Branching Ratio (BR), we would have to
revise the traditional $W$-anni\-hi\-la\-tion diagram\footnote{$^*$}
{By this we mean the mechanism by which the $c \bar s $ components of the
$D_s^+$ annihilate into a virtual $W^+$ which then decays into a pair of
on shell quarks which fragment into the final hadrons. Such a diagram is
depressed by the helicity conserving $W$-coupling to quarks.}
and other mechanisms of $W$-annihilation would have to be invoked to explain
some decays that would otherwise be inexplicable. One such possibility
will be briefly discussed.

\noindent
Although the considerations that follow are largely qualitative, we
believe that our reasoning is tight enough to make the ensuing scheme
self-contained and worth being experimentally investigated and tested.

Recently, an exhaustive experimental analysis of the 3-pion final state
decays of the charm pseudoscalar $D_s^+$ meson has been offered [1] which
increases significantly the existing statistics and consequently sheds
considerable light in a very confused picture. In spite of the (still
sizeable) error bars, this Dalitz plot analysis modifies considerably
the pre-existing situation [2]. The most important conclusion reached
in [1] concerns the smallness (in fact, the insignificance) of the {\it
non-resonating} $D_s^+ \rightarrow 3 \pi$ channel credited previously
with a non insignificant BR = ($1.01 \pm 0.35$)\% [2]. In addition, in
this analysis, a) two new channels, $D_s^+ \rightarrow f_2(1270) \,
\pi^+$ and $D_s^+ \to f_0(1300) \, \pi^+$ have been discovered
(admittedly though, the error bars are still quite large), b) the
absence of a significant channel $D_s^+ \rightarrow \rho^0\, \pi^+$
has been confirmed and, finally, c) the existence of a sizeable BR in
the (already observed) channel $D_s^+ \rightarrow f_0(980) \, \pi^+$ has
been substantiated.

While the observation of a non-resonant three-pion decay would have
represented a clear signature in favour of a non-spectator $D_s^+$
decay, its smallness (compatible with zero [1]), unfortunately, does
not convey an equally unambiguous (albeit negative) result. It is, in
fact, well known that a relevant component of three-body decays of
charmed particles comes from two-body decays in which one of the decay
products is in turn a resonant state. Thus, the fact that the
non-resonant $D_s^+ \rightarrow 3 \pi$ channel is very small, confirms
simply something that was already known; differently stated, the
smallness of a non-resonant three-pion decay channel by itself does
not imply the absence of $W$-annihilation type diagrams in $D_s^+$ decay.

Quite analogously, the observation of the decay $D_s^+ \rightarrow \rho^0
\, \pi^+$ would have been a clear signature of a non-spectator diagram.
In point of fact, based on the experimental observation of a significant
BR for the decay $D^0 \rightarrow \bar K^0 \, \phi$ [$(0.83 \pm 0.12)
\times 10^{-2}$] [2]
which denotes a large non-spectator component in $D^0$ decay, some
theoretical models [3, 4] had predicted a large BR for the decay
$D_s^+ \rightarrow \rho^0\, \pi^+$. This prediction was not met by the
data [2] for which the upper limit is
BR[$D_s^+ \rightarrow \rho^0\, \pi^+$] $\lsim 2.8 \times 10^{-3}.$

Again, however, the absence of this channel may have other explanations
and does not conclusively exclude the contribution of non-spectator
diagrams. To provide an example of how this may come about, let us point
out that a possible explanation for such a constraining experimental
result could be found in the very form of the isospin component
$\rho^0$ wave function which is $1/{\sqrt 2} \times (u \bar u - d \bar d)$.
If, in fact, we assume {\it factorization} of the $\rho^0\pi^+$ wave
function
$$ \eqalign{
\langle \rho^0\, \pi^+ | H_W | D_s^+ \rangle &= {1 \over \sqrt 2} \,
\langle (u \bar u - d \bar d) \, (u \bar d) | H_W | D_s^+ \rangle \cr
&= {1 \over \sqrt 2} \, \left[ \langle (u \bar u)\,
(u \bar d)| H_W | D_s^+ \rangle - \langle (d \bar d)
(u \bar d) | H_W | D_s^+ \rangle \right] \approx 0 \cr}
\eqno(1) $$
we are led to an immediate explanation for the smallness of this
particular decay channel.

This possibility, (which appears extremely natural and had not been
taken into account in the afore-mentioned estimates), can easily
be checked since it also predicts the absence
of other decay channels involving the direct ({\it i.e. non-resonant})
production of either one $\pi^0$ or of one $\rho^0$ when a
factorization analogous to (1) can be performed. Care must be taken, however,
be taken; this {\it no-go} mechanism does not apply when these particles
($\rho^0$ or $\pi^0$) can only be produced from either one, but {\it
not from both} the two $u \bar u$ and $d \bar d$ configurations. Thus,
in order to check experimentally that this mechanism is indeed
responsible for the absence of the $\rho^0 \pi^+$ final state, we suggest,
as an example, that the channel $D_s^+ \rightarrow \phi \, \pi^+ \,
\pi^0$ be investigated. If our conjecture is correct, the BR for this
reaction should also be essentially zero. As a matter of fact, this
decay channel has not been observed so far.

Assuming now that the previous factorization of the isospin wave
function is indeed responsible for the strong suppression of the decay
channel $D_s^+ \rightarrow \rho^0 \, \pi^+$, an immediate test of its
validity and, at the same time, of the existence of $W$-annihilation
diagrams can be suggested. The isospin wave function of the
$\omega (782)$ being $1/{\sqrt 2} \times (u \bar u + d \bar d)$, the
decay $D_s^+ \rightarrow \omega \, \pi^+$ (which can only proceed via
$W$-annihilation) should {\it not} be suppressed by the mechanism which
we have invoked for the case $D_s^+ \rightarrow \rho^0\, \pi^+$.

A (very rough) evaluation of the BR[$D_s^+ \rightarrow \omega (782)\,
\pi^+$] can be given by the same argument used originally [4] to
estimate $D_s^+ \rightarrow \rho^0\, \pi^+$. The form factor involved
in the latter is dominated by the $a_1(1260)$ pole while the form factor
involved in the decay $D_s^+ \rightarrow \omega (782) \, \pi^+$ is
dominated by the $b_1(1235)$ pole. Assuming the various parameters to be
comparable and the normalization constant to be of order unity
[3, 5], we can expect
$$ {\rm BR} \, [D_s^+ \rightarrow \omega (782)\, \pi^+] \simeq 1\%
\eqno(2)$$
if our assumption is correct. The experimental data do not provide at
the moment any definite answer about the existence of the above decay
channel; the present situation offers only a rather large upper bound,
BR[$D_s^+ \to \omega(782) \pi^+] < 1.7\%$ [2]. A careful experimental
search of this channel is highly necessary since it appears to be a
crucial test to prove or disprove the existence of non-spectator
contributions in $D^+_s$ decay.

The observation, with a BR of the order of 1\% as suggested in Eq. (2),
of the $D_s^+ \rightarrow \omega (782)\, \pi^+$ decay mode, would be
a strong argument in favour both of a sizeable $W$-annihilation
contribution and of the wave function factorization argument, Eq. (1).
In such a case other decay channels which can only (or mainly) proceed
through $W$-annihilation should be observed with comparable branching
ratios. On the other hand, the non observation of this decay (or a
very tiny BR $\ll 1\%$), would definitely exclude a relevant
contribution from the usual $W$-annihilation decay mechanism;
in this case, other mechanisms should be responsible for the
observation of decays which could have occurred via simple
$W$-annihilation; one such possibility will be discussed shortly.

Let us recall once again that the new data [1] indicate that the decay
$D_s^+ \to 3\pi$ proceeds mainly through the resonant channels
$D_s^+ \to R \, \pi^+$ (with $R = f_0(980)$, $f_0(1300)$ and
$f_2(1270)$).
According to the quark model, these resonances are classified as
isospin singlets with different mixtures of $u\bar u$, $d\bar d$ and
$s\bar s$ components, whose exact nature is still much debated [6]. A
precise knowledge of their quark content is of the greatest importance
to understand the relevance of non-spectator contributions in
the decay $D_s^+ \to 3\pi$. The presence of a dominant $s\bar s$
component (as in the case of the $f_0(980)$), suggests that the
$D_s^+ \to f_0(980) \pi^+$ decay should occur through a spectator
process; on the other hand a negligible $s\bar s$ component (which
seems to be the case for $f_2(1270)$ and $f_0(1300)$), would rather
point to a non-spectator decay.

Let us consider separately the $f_0(980)$ and the $f_2(1270)$,
$f_0(1300)$ resonances. The first one, which contributes the largest
part of the $D_s^+ \to 3\pi$ decay rate, is well known to have a large
branching ratio into $K\bar K$ strange mesons, BR$[f_0(980) \to
K\bar K = (21.9 \pm 2.4)\%]$ [2]. Considering the tiny phase space
available for such a reaction it is natural to expect the $f_0(980)$
to be dominated by a $s\bar s$ quark component. As a consequence, the
decay $D_s^+ \to f_0(980) \pi^+$ is most certainly occurring via a
dominant spectator diagram.

For the other two resonances which also have been observed in the
$D_s^+ \to 3\pi$ channel the situation is quite different: despite a
larger phase space available, their branching ratios for decays into
strange mesons are much smaller [2]: BR[$f_2(1270)$ $\to K\bar K =
(4.6 \pm 0.5)\%]$ and BR[$f_0(1300) \to K\bar K = (7.5 \pm 0.9)\%]$.
Both values are similar to those observed for the decays into strange
mesons of particles whose $s\bar s$ component is known to be essentially
zero, like $a_2(1320)$, $\pi_2(1670)$ and $\rho_3(1690)$, for which,
typically, BR $\sim 5\%$. This strongly suggests that the $s\bar s$
content of $f_2(1270)$ and $f_0(1300)$ is negligible. The same
conclusion about the strange component of $f_0(980)$ and $f_0(1300)$
has been reached in Ref. [7].

A dominant $u\bar u$ and $d\bar d$ component of the $f_2(1270)$ and the
$f_0(1300)$ implies that the observed decays $D_s^+ \to f_2(1270)
\pi^+$ and $D_s^+ \to f_0(1300) \pi^+$ cannot occur through the
spectator diagram of $W$-radiation, but should rather proceed via the
non-spectator diagram of $W$-annihilation. As already discussed, the
relevance of this mechanism would definitely be confirmed by the
experimental observation of the decay $D_s^+ \to \omega(782) \pi^+$;
a branching ratio of the order of magnitude predicted by Eq. (2), would
be a clear signature in favour of the significance of the
$W$-annihilation mechanism and the usual non-spectator models could
explain all of these three decays.

In the opposite case -- the non observation of a sizeable branching
ratio for the $D_s^+ \to \omega(782) \pi^+$ decay -- there is no reason
to expect that if the $W$-annihilation mechanism does not work for the
$\omega$ vector meson it should work for the scalar or tensor ones, but
one should rather accept the usual helicity argument which forbids
simple $W$-annihilation. In this case, one would have to look for
different decay diagrams to account for the $D_s^+ \to f_2(1270) \pi^+$
and $D_s^+ \to f_0(1300) \pi^+$ modes. These new diagrams should
allow the production of scalar and tensor particles while
forbidding that of vector ones (like $\rho^0$ and $\omega$).

An interesting possibility is shown in Fig. 1, according to which the
$c$ and $\bar s$ quarks in the $D_s^+$ not only annihilate into a
virtual $W^+$, which directly generates the $\pi^+$,
but also into two gluons;
this avoids the helicity argument [8, 9]. For such a contribution to
be significant the two gluons must couple directly to a large gluon
component of the final meson: $f_0$ and $f_2$ have the right quantum
numbers and could indeed be produced via this scheme, but a $1^{--}$
vector meson state could not be produced owing to the positive
$C$-parity of a two gluon state.

This mechanism is similar to the one suggested previously to explain
the decays $J/\psi \to \rho \pi$ [10] and $\eta_c \to VV$ [11] (where
$V$ is a vector meson). These processes have in fact posed a challenge
in their theoretical interpretation; although forbidden by helicity
conservation, they have been known to occur for a long time and are
observed with significant branching ratios of order 1\% [2]. The reason
why the helicity conservation argument is overcome by the higher order
diagram of Fig. 1 lies in the fact that this diagram enhances the production
of a resonant $\pi^+ \pi^-$ state with an invariant mass $M$ via a Breit-Wigner
factor
$$ {1\over (M-M_G)^2 + \Gamma_G^2/4} \,, \eqno(3)$$
where $M_G$ and $\Gamma_G$ are respectively the mass and the width of
the gluonic state mixed with the $f_2(1270)$ or the $f_0(1300)$.

While, to the best of our knowledge, no other mechanisms have been
successful in predicting a significant BR for the reactions
$J/\psi \to \rho \pi$ and $\eta_c \to VV$ (explaining, at the same time,
the non observation of the $\psi^\prime \to \rho \pi$ decay)
the final word on the validity of the mechanism under discussion [10, 11]
has still to be said.
The intriguing possibility that a similar mechanism may be at the root
of the production of $f_2(1270)$ and $f_0(1300)$ in $D_s^+$ decay
leaves us one more chance of explaining these reactions, should
indeed the direct $W$-annihilation be proved absent {\it i.e.} should the
search of a significant $D_s^+ \to \omega(782) \pi^+$ decay be
unsuccessful. The mechanism of Fig. 1, however, demands a large coupling
to gluons of both $f_2(1270)$ and $f_0(1300)$ which at this stage of our
knowledge is neither excluded nor suggested by the data; this in itself
raises questions that deserve a careful investigation.

A detailed study of the $D_s^+$ decay modes will no doubt help in
clarifying a very complex situation; here we have just considered in a
somewhat limited and qualitative way the r\^ole of the annihilation
$c \bar s \rightarrow W^+$ in non-spectator decays of $D_s^+$ and we have
raised a number of questions and suggested a number of experimental
tests and investigations which could provide a much better insight into
these problems. Some conclusions can be safely reached:
the observation of the $D_s^+ \to \omega \pi^+$ decay mode with a
branching ratio $\sim$ 1\% would definitely prove the relevance of
the traditional simple $W$-annihilation contribution; the observation
of a small BR[$D_s^+ \to \omega \pi^+] \ll 1\%$ instead, would
indicate a negligible $W$-annihilation contribution and would pose the
problem of explaining some other decays which are so far believed to
occur through $c \bar s$ annihilation. Alternative mechanisms can be at
work in this case and we have suggested one. The solution of these
problems opens new interesting possibilities which we hope will soon
be investigated experimentally.

\vskip 24 pt
\noindent
{\bf Acknowledgements}
\vskip 6pt
\noindent
One of us (I.B.) would like to thank the INFN
for financial support; we are grateful to E. Levin, L. Moroni and
N. Nikolaev for useful discussions.

\vfill\eject
\noindent
{\bf References}
\vskip 12pt
\item {[ 1]} L. Moroni, {\it ``Hadronic Decays of Charm Mesons"}, presented
      in Lafex International School on High Energy Physics, LISHEP-95.
\item {[ 2]} PDG, {\it Phys. Rev.} {\bf D50} (1994) 1173
\item {[ 3]} U. Baur, A.J. Buras, J.-M. G\'erard and R. R\"uckl,
      {\it Phys. Lett.} {\bf 175B} (1986) 377
\item {[ 4]} I. Bediaga and E. Predazzi, {\it Phys. Lett.} {\bf 199B} (1987)
131
\item {[ 5]} D. Fakirov and B. Stech, {\it Nucl. Phys.} {\bf 133B} (1978) 315
\item {[ 6]} N.A. T\"ornqvist, preprint HU-SEFT R 1995-05
\item {[ 7]} M. Gourdin, Y.Y. Keum and X.Y. Pham, preprint PAR/LPTHE/95-09,
             hep-ph/9503326
\item {[ 8]} H. Fritsch and P. Minkowski, {\it Phys. Rep.} {\bf 73} (1981) 67
\item {[ 9]} M. Bander, D. Silverman and A. Soni, {\it Phys. Rev. Lett.}
             {\bf 44} (1980) 7; {\it Errata} {\bf 44} (1980) 962
\item {[10]} S.J. Brodsky, G.P. Lepage and S.F. Tuan, {\it Phys. Rev. Lett.}
             {\bf 59} (1987) 621
\item {[11]} M. Anselmino, M. Genovese and E. Predazzi, {\it Phys. Rev.}
             {\bf D44} (1991) 1597
\bye